%% file: main.tex
\documentclass[10pt,conference]{IEEEtran}
\usepackage{cite}
\usepackage{amsmath,amssymb,amsfonts}
\usepackage{algorithmic}
\usepackage{graphicx}
\usepackage{textcomp}
\usepackage{fancyhdr}

\usepackage{threeparttable}
\usepackage[table,dvipsnames]{xcolor}
\usepackage{tabularx,ragged2e,makecell}
\usepackage{array}
\usepackage{pifont} 
\usepackage{soul}
\usepackage[normalem]{ulem}
\usepackage{babel}
\usepackage{comment}
\usepackage{caption}
\usepackage{subcaption}
\usepackage{multicol}
\usepackage{multirow}
\usepackage{tikz}
\usetikzlibrary{arrows.meta, positioning}
\usepackage{enumitem}
\usepackage{booktabs}
\usepackage[bookmarks=true,breaklinks=true,letterpaper=true,colorlinks,citecolor=blue,linkcolor=blue,urlcolor=blue]{hyperref}
\usepackage{cleveref}

\newcommand{\namef}{HARMONI}
\newcommand{\nname}{Sangam}
\newcommand{\hpcayear}{2026}
\newcommand{\circnum}[1]{%
  \tikz[baseline=(char.base)]\node[shape=circle, draw, fill=black, text=white, 
    inner sep=0.6pt, minimum size=1em, font=\footnotesize] (char) {#1};%
}

\title{Sangam: A Chiplet-Based DRAM-PIM Accelerator with CXL Integration for LLM Inferencing}

\newcommand\hpcaauthors{\\Khyati Kiyawat$^*\dagger$, Zhenxing Fan$^*\dagger$, Yasas Seneviratne$\dagger$, Morteza Baradaran$\dagger$,\\ Akhil Shekar$\dagger$, Zihan Xia$\ddagger$, Mingu Kang$\ddagger$, and Kevin Skadron$\dagger$}
\newcommand\hpcaaffiliation{University of Virginia$\dagger$, University of California, San Diego$\ddagger$\\$^*$Equal Contribution}

\author{
  \ifdefined\hpcacameraready
    
  \else
    \IEEEauthorblockN{\hpcaauthors{}}
      \IEEEauthorblockA{
        \hpcaaffiliation{} 
    }
  \fi 
}

\fancypagestyle{camerareadyfirstpage}{%
  \fancyhead{}
  
  \fancyhead[C]{
    \ifdefined\aeopen
    \parbox[][12mm][t]{13.5cm}{\hpcayear{} IEEE International Symposium on High-Performance Computer Architecture (HPCA)}    
    \else
      \ifdefined\aereviewed
      \parbox[][12mm][t]{13.5cm}{\hpcayear{} IEEE International Symposium on High-Performance Computer Architecture (HPCA)}
      \else
      \ifdefined\aereproduced
      \parbox[][12mm][t]{13.5cm}{\hpcayear{} IEEE International Symposium on High-Performance Computer Architecture (HPCA)}
      \else
      \parbox[][0mm][t]{13.5cm}{\hpcayear{} IEEE International Symposium on High-Performance Computer Architecture (HPCA)}
    \fi 
    \fi 
    \fi 
    \ifdefined\aeopen 
      \includegraphics[width=12mm,height=12mm]{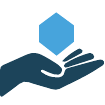}
    \fi 
    \ifdefined\aereviewed
      \includegraphics[width=12mm,height=12mm]{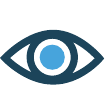}
    \fi 
  }
  \fancyfoot[C]{}
}
\newpage

\begin{document}

\thispagestyle{plain}  
\newpage  

\maketitle

\ifdefined\hpcacameraready 
  \thispagestyle{camerareadyfirstpage}
  \pagestyle{empty}
\else
  \thispagestyle{plain}
  \pagestyle{plain}
\fi

\newcommand{\hpcaheight}{0mm}
\ifdefined\eaopen
\renewcommand{\hpcaheight}{12mm}
\fi


\input{0-abs}

\input{1-introduction}

\input{2-background}

\input{4-architecture}

\input{5-methodology}

\input{6-results}

\input{7-discussion}

\input{8-conclusion}

\input{9-acknowledgement}


\bibliographystyle{IEEEtran}
\bibliography{refs}

\end{document}

%% file: 0-abs.tex
\begin{abstract}
Large Language Models (LLMs) are becoming increasingly data-intensive due to  growing model sizes, and they are becoming memory-bound as the context length and, consequently, the key-value (KV) cache size increase. Inference, particularly the decoding phase, is dominated by memory-bound GEMV or flat GEMM operations with low operational intensity (OI), making it well-suited for processing-in-memory (PIM) approaches. However, existing in/near-memory solutions face critical limitations such as reduced memory capacity due to the high area cost of integrating processing elements (PEs) within DRAM chips, and limited PE capability due to the constraints of DRAM fabrication technology. This work presents a chiplet-based memory module that addresses these limitations by decoupling logic and memory into chiplets fabricated in heterogeneous technology nodes and connected via an interposer. 
The logic chiplets sustain high bandwidth access to the DRAM chiplets, which house the memory banks, and enable the integration of advanced processing components such as systolic arrays and SRAM-based buffers to accelerate memory-bound GEMM kernels, capabilities that were not feasible in prior PIM architectures. We propose Sangam, a CXL-attached PIM-chiplet based memory module that can either act as a drop-in replacement for GPUs or co-executes along side the GPUs. Sangam achieves speedup of  3.93, 4.22, 2.82x speedup in end-to-end query latency, 10.3, 9.5, 6.36x greater decoding throughput, and order of magnitude energy savings compared to an H100 GPU for varying input size, output length, and batch size on LLaMA 2-7B, Mistral-7B, and LLaMA 3-70B, respectively.
\end{abstract}

%% file: 1-introduction.tex
\section{Introduction} \label{sec:Introduction}

The global AI inference market is growing rapidly and 
generative AI inference  is estimated to drive over 75\% of AI compute and energy demand in U.S. data centers \cite{inference_dominance}. GenAI power demand in the US is projected to reach 800 TWh by 2030, increasing from 8TWh in 2024 \cite{genAIEnergy}.  

Almost all leading large language models (LLMs), vision/speech models, and multimodal systems are based on the Transformer architecture \cite{attentionIsAllYouNeed}, with applications ranging from summarization and chatbots\cite{noauthor_claude_nodate} to language-translation \cite{chung_scaling_2022} and code generation\cite{noauthor_github_2025, thakur_benchmarking_2022}, among others. 
However, current state-of-the-art inference platforms rely heavily on costly, high-end servers equipped with multiple GPUs and high-bandwidth memory (HBM). 
Such hardware configurations are capital-intensive for datacenters and expensive to rent. An example is the NVIDIA's H100 GPU which costs around \$25k-30k and the rental costs for the same are estimated at \$3 per hour \cite{GPUcost}, not to mention the constrained availability of such hardware.

\begin{figure*}[t]
    \centering
    \includegraphics[width=0.9\textwidth]{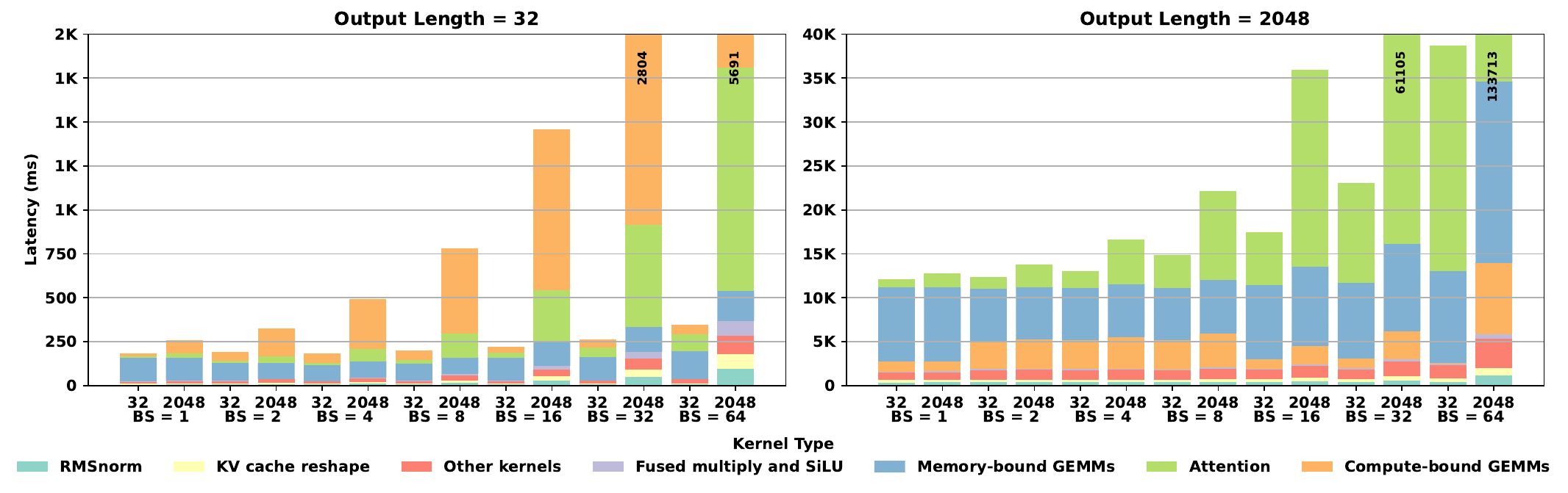}
    \caption{LLaMA 2-7B  inference kernel latency breakdown on one H100 GPU. Left and right figures show breakdown under varying batch sizes and input lengths when output length is 32 and 2048, respectively. (BS = Batch Size.)}
    \label{fig:GPU_latency_breakdown}
\end{figure*}

Prior works \cite {recasens_mind_2025,kundu_performance_2024} find that, during the LLM token generation phase (predicting one output token at a time), inference is DRAM-bandwidth bound in most scenarios, and even prefill (processing the input prompt) is sometimes bandwidth bound (see \Cref{subsec:GEMM_Shapes_and_Ai}).
For example, GPU utilization during single-batch LLM inference workloads often remains below 20\% \cite{mindTheMemoryGap, attacc, CXL-PNM} and never more than 40\% as reported by \cite{Neupim}, indicating a substantial mismatch between provisioned compute and effective throughput, leading to poor utilization of an expensive resource such as the GPU. 
Techniques such as batching are employed to reduce memory accesses and improve processor utilization by enabling the reuse of weight matrices during GEMM operations. However, the achievable batch size is limited by memory capacity \cite{attacc}, since the KV cache grows linearly with both batch size and sequence length, and by service-level objectives (SLOs) that impose minimum latency requirements, thereby restricting the extent of batching \cite{PAPI}.

The inefficiency in state-of-the-art inference platforms primarily arises from memory bottlenecks, which manifest in two key dimensions: capacity and bandwidth. The escalating scale of GenAI inference underscores the need for a more balanced and cost-efficient design paradigm. Addressing this demand calls for research focused on two critical \textbf{C}hallenges: \textbf{(C1)} developing systems with scalable memory capacity, and \textbf{(C2)} minimizing over-provisioning of costly, underutilized hardware.
One promising direction is to extend the functionality of an inference platform's memory by augmenting it with processing-in-memory (PIM) capabilities that provide higher throughput for memory-bound computations by leveraging the very high bandwidth internal to the DRAM architecture. 
A compute-enabled memory system can potentially replace a GPU for much lower cost, or alternatively, enhance the GPU for a modest extra cost. For these reasons, Davies et al.~\cite{kozyrakis2025efficient} argue that DRAM-based inference solutions provide better capacity and bandwidth to achieve superior tokens/Watt and tokens/\$ performance.


Indeed, prior works \cite{Aquabolt, Newton,  Neupim, attacc, PAPI, CENT, pim-dl, pim-ai, pim-gpt, duplex, transpim, sal-pim, pipepim, specpim, stepstone} have demonstrated PIM as a potential solution to accelerate AI inference due to high internal bandwidth (orders of magnitude higher than a GPU with HBM at iso-capacity). However, most PIM approaches are limited in compute capability due to area and DRAM fabrication constraints. Thus, they targetted kernels with very low operational intensity (OI) of approximately 1 FLOP/byte. Supporting kernels with higher operational intensity (OI), such as matrix multiplication for prefill or batched decoding, requires more expensive logic units along with some buffering capacity to enable data reuse. 
Unfortunately, {\textbf{(C3.1)} adding logic units
and buffers near the memory cell array inside the DRAM die sacrifices memory capacity (reducing the memory density by 25\%-75\% \cite{AiM, PAPI, UPMEM-AI, Aquabolt}). 
Furthermore, \Cref{fig:GPU_latency_breakdown} provides latency breakdown of each kernel contributing to inference as a function of batch size and input sequence length, with two example output lengths: 32 and 2048. When output length is 32 (\Cref{fig:GPU_latency_breakdown}), memory-bound GEMM kernels are the main contributor only when input is also small (32), while for larger batch sizes and larger input size (2048), both attention and compute-bound GEMM dominate inference latency. Notably, when the output length is 2048 (right figure), memory-bound GEMMs consistently account for a large share of latency across all batch sizes. This shows that end-to-end LLM inference relies heavily on GEMM kernels to support input processing and batched processing, 
{\em with kernels computing \textgreater 1} but {\textless 200 FLOPs/byte} contributing the most to the end to end latency (\Cref{subsec:GEMM_Shapes_and_Ai}). These {\em flat-GEMMs} are memory-bound on GPUs, but inefficient when unrolled to a sequence of GEMV (matrix-vector) operations in DRAM PIM, which loses opportunities for data reuse through tiling, e.g. via systolic arrays. \textbf{(C3.2)} In addition, many prior PIM works~\cite{Aquabolt, UPMEM-AI, AiM} that put logic at the bank-level requires inefficient and expensive bank-to-bank data transfers. 

Adding any logic transistors in a DRAM process occupies 10$\times$ more area and runs at a slower clock compared to a logic process \cite{upmem-hotchips}. However, fabricating the compute units using logic process technology require heterogeneous integration of DRAM and logic chiplets. Recent advances in 2.5D and 3D integration have explored alternative pathways to couple DRAM with compute logic. For example, 3D stacked DRAM+logic designs using wafer-on-wafer (WoW) or chip-on-wafer (CoW) bonding place functional units on a logic base die, enabling compute closer to memory~\cite{zhang_challenges_2022,lu_tsv-based_2017, ohba_review_2022}. \textbf{(C3.3)} While these approaches benefit from high bandwidth within a stack, the  interface to the logic layer remains constrained by the pitch of through-silicon vias (TSVs), limiting  bandwidth to the compute units, which must be shared across banks/dies. For example, DUPLEX~\cite{duplex} only increases TSV density by 4X, which is insufficient for all-bank PIM.

To address the above challenges we present Sangam
\footnote{``Sangam" is a Sanskrit word meaning confluence. In the context of this work, it signifies the integration of memory and compute chiplets.}, 
a novel system architecture that integrates chiplet-based DDR-PIM accelerator modules through CXL interconnects. These inference-only modules can either replace the GPU or complement it by offloading large GEMMs to GPU.

 
\textbf{Chiplet Integration for DRAM PIM.} \textcolor{ForestGreen}{} Xia et al. \cite{ChipletTVLSI} proposed a chiplet-based DRAM design that partitions a conventional DRAM chip into multiple memory chiplets, keeping the banks in DRAM chiplets while moving the center stripe containing the column decoder, bank-interface buffer, and control circuitry to a logic chiplet, as illustrated in \Cref{fig:Sangam_system}b. Our design \textbf{S}olution \textbf{S1} proposes to extend and augment the center stripe chiplet with compute units and SRAM buffer designed for LLM kernels' execution and communication among the chiplets which addresses the key memory density challenge \textbf{C3.1}. By directly attaching bank-level PIM units to all banks across the DRAM chiplets, Sangam is, to our knowledge, the first architecture that exposes the full bank-level bandwidth of commodity DRAM to PIM functional units fabricated in a logic process, addressing the challenge in alternative integration solutions \textbf{C3.3}. The center stripe have access to data from all DRAM banks in a chip, enabling high aggregate bandwidth (up to 400 GBps) to the compute units without needing expensive bank-to-bank interconnects, thus solving the bank-to-bank communication challenge \textbf{C3.2}. We implement systolic arrays in the logic chiplet without impacting DRAM capacity. This allow GEMMs across a range of sizes to be supported efficiently. 
\textbf{CXL Interface.} Sangam communicates with the host processor and storage via a CXL switch interface \textbf{(S2)}
enabling direct loading of the inference requests from the host CPU and also Large Language Model (LLM) weights into Sangam modules from the storage. This approach allows high capacity while alleviating the strain on the CPU's chip area to accommodate sufficient pins to support a large number of memory channels, because addition of each memory channel is costly in area due to the expensive DDR interface. This also frees the host processor to focus on execution of non-LLM workloads tasks using standard/conventional DIMMs. The DDR-based Sangam modules offer higher memory capacity (GB/\$) than GDDR- and HBM-based PIM approaches, enabling support for larger models and addressing the critical challenge \textbf{(C1)} of building systems with scalable memory capacity.

Although GPUs outperform Sangam for large GEMMs, especially the projections in prefill with large inputs, Sangam excels across a range of GEMM sizes, and  provides better end-to-end performance unless the input is very large and the output is small. 
This behavior makes Sangam well suited for scenarios such as reasoning models with short prompts and long outputs, multi-user or multi-turn conversations with prefix caching, and decode-only workloads that leverage stored KV caches.
The proposed solution addresses \textbf{S3} by ensuring that users pay only for the smart memory (Sangam modules) needed for their target LLMs, eliminating the need for additional expensive accelerators that are often underutilized, which directly tackles the critical challenge \textbf{C2}. Details of the system design and Sangam architecture are discussed in \Cref{sec:Architecture}.

Key contributions include: 
\begin{itemize}
    \item To the best of our knowledge, Sangam is the first work that introduces the benefits of chiplet integration for DRAM-PIM to increase compute capability  of PIM architectures, enabling use of systolic arrays in PIM. 
    \item Sangam's CXL attached PIM is: 1) more scalable for large LLMs  compared to conventional DIMM, since it does not need memory controller/bump area on the CPU, and  
    2) can be attached to GPU if need be.
    \item We propose a new data placement aware hierarchical PIM architecture evaluation framework for genAI workloads ``HARMONI" - \textbf{H}ierarchical \textbf{AR}chitecture \textbf{MO}deling for \textbf{N}ear/\textbf{I}n Memory Computing.
    Sangam achieves speedup of  3.93, 4.22, 2.82x speedup in end-to-end query latency, 10.3, 9.5, 6.36x greater decoding throughput, and substantial energy savings compared to an H100 GPU across various input, output, and batch sizes on LLaMA 2-7B, Mistral-7B, and LLaMA 3-70B, respectively.

\end{itemize}

%% file: 2-background.tex
\section{Background} \label{sec:Background}
\subsection{LLM Inference}
\label{subsec:LLM_execution_flow}
The inference process comprises two phases: Prefill (Summarization) and Decode (Generation). During prefill, the input prompt is tokenized into a sequence of token IDs. Each ID undergoes an embedding lookup, retrieving its corresponding high-dimensional embedding vector from the model's vocabulary, creating an input matrix. This embedding matrix (X) is processed through a stack of identical decoder layers sequentially. Each layer incrementally modifies X and the final X is passed to a Language Modeling Head (LM Head) that generates logits (unnormalizded probabilities) for each input token. These logits are passed through a softmax function to get a probability distribution over the model vocabulary. The next token is selected from this distribution. In the decode phase, subsequent tokens are generated iteratively, one at a time, using the same decoder stack. Here, the input is an embedding vector\footnote{This explanation is considering only one input prompt (batch=1) for simplicity} rather than a full matrix. Crucially, the prefill phase builds the Key-Value (KV) cache for all input tokens, while the decode phase loads and appends this cache with each new token (during attention). The KV cache is unique to each decoder layer and inference request and grows sequentially with the number of tokens making the decode phase both memory capacity and bandwidth bound. KV cache optimization techniques such as compression, pruning, etc \cite{KVsurvey} delay this bottleneck but do not eliminate it.

Each decoder layer consists of sub-layers: First, Attention: The embedding (X) per token is projected into three vectors:  Query (Q),  Key (K), and  Value (V). 
Key and Value from  past tokens is used (via the KV cache) to compute a score (followed by a softmax) and context. 
Second, Feed-Forward Network: The output from each head is concatenated and processed through output, up, gate, and down projections. 
Residual connections and Root-mean Square (RMS) normalization are applied around each of these two sub-layers. 

\subsection{GEMM Shapes and Operational Intensity (OI)}
\label{subsec:GEMM_Shapes_and_Ai}

\begin{table}[t]
\centering
\caption{GEMM dimensions and operational intensity (OI) for LLaMA 2-7B in prefill and decode phases.}
\begin{threeparttable}
\begin{tabular}{lcccc}
\toprule
\textbf{GEMM Kernel} & \textbf{M} & \textbf{K} & \textbf{N} & \textbf{OI} \\
\midrule
\multicolumn{5}{l}{\textbf{Prefill Phase}} \\
QKV Projection             & B$\cdot$I & 4096 & 3$\cdot$4096 & 768 \\
Score (Q × K$^\top$)   & I & 128 & I & 43 \\
Context (Softmax(Score) × V)    & I & I & 128 & 43 \\
Output Projection         & B$\cdot$I & 4096 & 4096 & 683 \\
Gate/Up Projection         & B$\cdot$I & 4096 & 11008 & 762 \\
Down Projection            & B$\cdot$I & 11008 & 4096 & 762 \\
LM Head                    & B$\cdot$I & 4096 & 32000 & 799 \\
\midrule
\multicolumn{5}{l}{\textbf{Decode Phase}} \\
QKV Projection             & B & 4096 & 3$\cdot$4096 & 8 \\
Score (Q × K$^\top$)   & 1 & 128 & (Past+1) & 1 \\
Context (Softmax(Score) × V)    & 1 & (Past+1) & 128 & 1 \\
Output Projection         & B & 4096 & 4096 & 8 \\
Gate/Up Projection         & B & 4096 & 11008 & 8 \\
Down Projection            & B & 11008 & 4096 & 8 \\
LM Head                    & B & 4096 & 32000 & 8 \\
\bottomrule
\end{tabular}
\begin{tablenotes}
\footnotesize
\item B: Batch size, I: Input length, Past: cached tokens in decode. 
LLaMA 2-7B uses 4096 hidden dim, 32 heads (128 dim each), 11008 FFN dim, and 32K vocab. OI is computed with B=8, I=128.
\end{tablenotes}
\end{threeparttable}
\label{tab:gemm_shapes_ai}
\end{table}

A typical transformer-based LLM inference process contains eight GEMM kernels (we treat attention GEMV as a special GEMM case): (1) QKV projection (QKV generation): embedding vectors are multiplied with the corresponding weight matrices; Attention computation: Q, K, and V matrices are multiplied to compute (2) attention scores and (3) context vectors; (4) Output projection; Feedforward (5) gate (6) up and (7) down projections (embedding vectors are multiplied with the feedforward gate, up and down matrices; (8) LM head: the final output embedding is multiplied with the language model head (vocabulary projection matrix) to produce logits.

\begin{figure}
    \centering
    \includegraphics[width=0.93\linewidth]{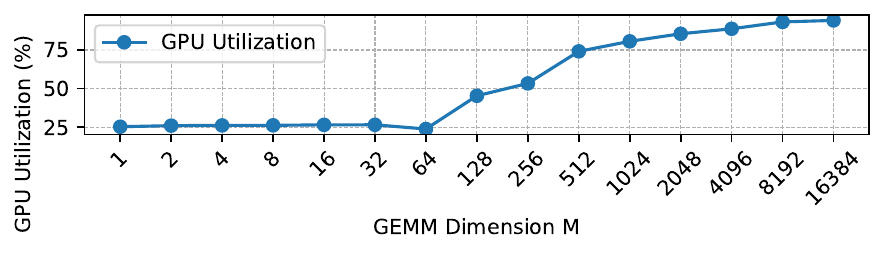}
    \caption{GPU utilization for GEMMs with varying M dimension.}
    \label{fig:gpu-utilization}
\end{figure}

We represent a GEMM operation as \texttt{GEMM}($M, K, N$), where a matrix $A \in \mathbb{R}^{M \times K}$ is multiplied by a matrix $B \in \mathbb{R}^{K \times N}$ to produce an output matrix $C \in \mathbb{R}^{M \times N}$. Table \ref{tab:gemm_shapes_ai} summarizes the GEMM kernels' shape that appear in the inference process of LLaMA 2-7B, along with their operational intensity (OI) when the batch size is 8 and the input sequence length is 128. The GEMM shapes, particularly the $M$ dimension, are determined by both the batch size and input length during the summarization phase, but only by the batch size during the decode phase. Since, the batch size on GPU is limited (27 \cite{attacc}) due to either SLO constraint or the HBM capacity, 
the GEMM kernels often have a small $M$ dimension but very large $K$ and $N$ dimensions.This produces a ``flat-shaped'' GEMM with low OI, which we refer to as a \textit{flat GEMM} \cite{hong_flashdecoding_2024}. An extreme case occurs when $M = 1$, reducing the operation to a GEMV. Limited by HBM bandwidth, a high-end GPU like H100 only achieves around 25\% of peak throughput when M is smaller than 128 as shown in \Cref{fig:gpu-utilization}. 

\subsection{Roofline analysis}
\label{subsec:roofline}
\begin{figure}
    \centering    \includegraphics[width=0.45\textwidth]{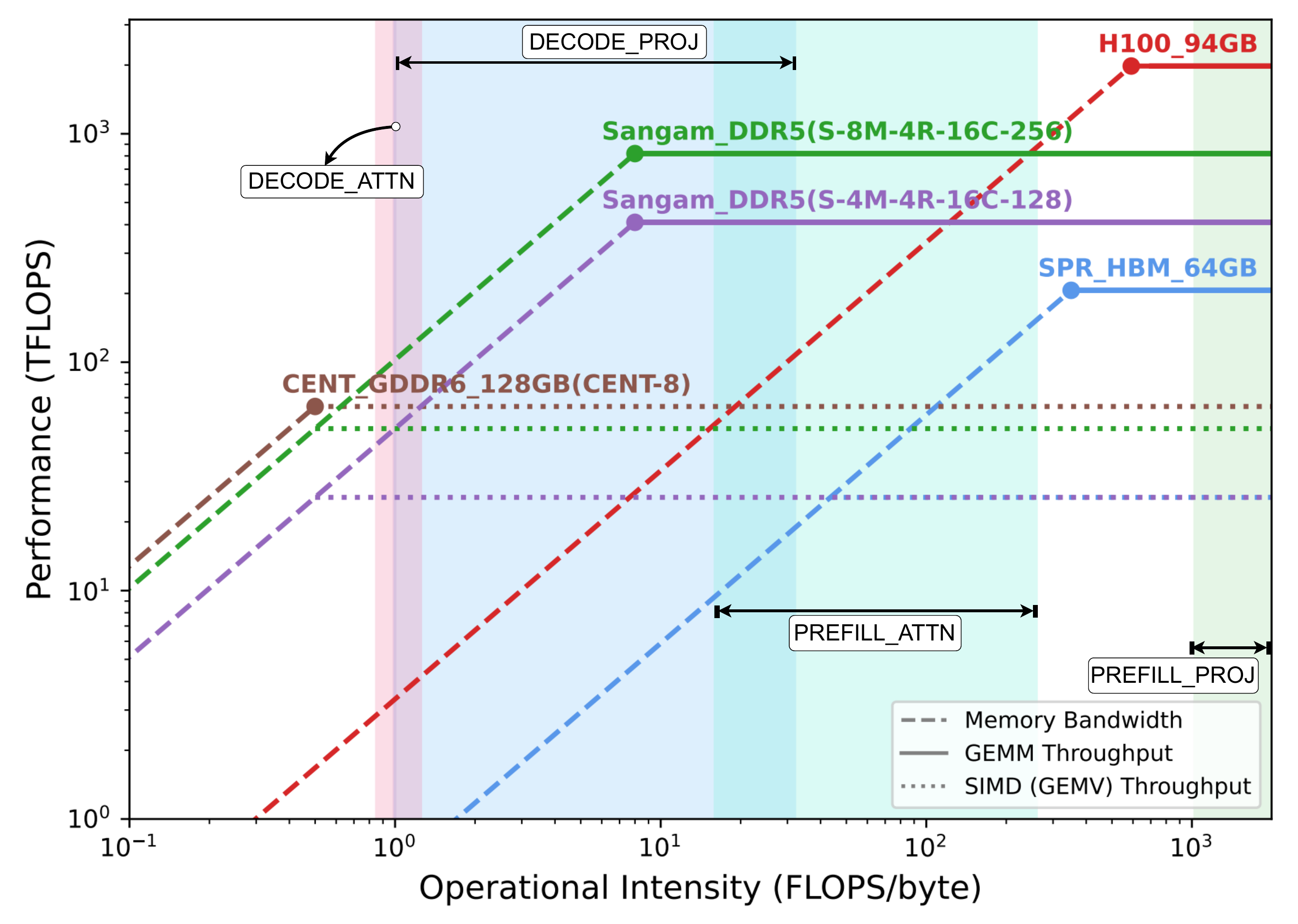}
    \caption{OI characterization for kernels in different phases of E2E LLM inference on different rooflines. Here, the OI is calculated for 2048 inputs, 2048 outputs, and varying batch size (in the range of 1-64) for LLaMA 2-7B model.}
\label{fig:System_Roofline_with_AI_characterization}
\end{figure}

\Cref{fig:System_Roofline_with_AI_characterization} shows the roofline plot for an H100 GPU \cite{noauthor_nvidia_nodate} with 80GB HBM, Intel Xeon Sapphire Rapids (SPR) with 64GB HBM \cite{saphire-rapids}, CENT \cite{CENT}, and Sangam system with different capacity configurations. CENT and Sangam have higher memory bandwidth compared to GPU and CPU. Even though at iso-capacity vs.\ CENT, Sangam (DDR5) has lower bandwidth than CENT (GDDR6), it has more compute throughput, thanks to chiplet integration and systolic arrays. The different colored vertical shades on the plot show the range of OI pertaining to different phases of LLM inference attention and projection GEMMs in prefill and decode. As discussed earlier, GEMMs can become memory-bandwidth bound on GPU, e.g. even for a large input size such as 2048 at different batch sizes; attention in the prefill phase with OI in the range of 15 to 250 FLOPs/byte is memory-bound on H100, making Sangam a suitable candidate. It is important to note that the projections in prefill are compute-intensive, showing H100 a clear winner. However, if we consider the prefill latency or time to first token (TTFT), it is a sum of both the projections and attention in the prefill phase. 

\subsection{Systolic arrays}
Systolic arrays~\cite{kung_why_1982} are a widely adopted architectural design for accelerating matrix multiplications. 
A systolic array consists of a grid of processing elements (PEs) connected in a regular pattern, where data flows  from one PE to the next. Each PE performs multiply–accumulate (MAC) operations, passing partial results along to its neighbors. This design enables high data reuse, reduces memory bandwidth demands, and achieves high throughput when matrices are well-matched to the array dimensions. Systolic arrays achieve their peak efficiency when executing sufficiently large and \emph{balanced} GEMMs, where the dimensions $M$, $K$, and $N$ are sufficiently large to keep all PEs busy and amortize data movement overhead. This is typically the case for training workloads or inference with large batch sizes. However, as discussed in Section~\ref{subsec:GEMM_Shapes_and_Ai}, LLM inference often produces GEMM shapes that are \emph{flat}---having a small $M$ dimension relative to large $K$ and $N$ dimensions. To address this challenge, our design adopts \emph{a set of small systolic arrays} (e.g., $8\times 8$) operating in parallel. This granularity better matches the $M$ dimension of flat GEMMs, allowing each array to be fully utilized even when $M$ is small. By instantiating multiple such arrays, our architecture preserves high parallelism across GEMM tiles while maintaining efficiency for both compute-bound and memory-bound flat GEMMs. We explain how we map a flat GEMM to the a systolic array in \Cref{subsec:Data Mapping}.

\subsection{Chiplet DRAM}
As monolithic integration increasingly struggles with yield loss and the rising cost of building large chips at advanced nodes, chiplet-based integration has emerged as a scalable and cost-effective alternative. By partitioning systems into smaller dies, each optimized for specific functions such as logic, I/O, or analog circuits, and connecting them via high-bandwidth die-to-die (D2D) interconnects (e.g., UCIe\cite{UCIeSpec1.1}, supporting up to 64~GB/s per lane), chiplets enable improved modularity, better yield, and enhanced process specialization.

\begin{figure}
    \centering
    \includegraphics[width=0.93\linewidth]{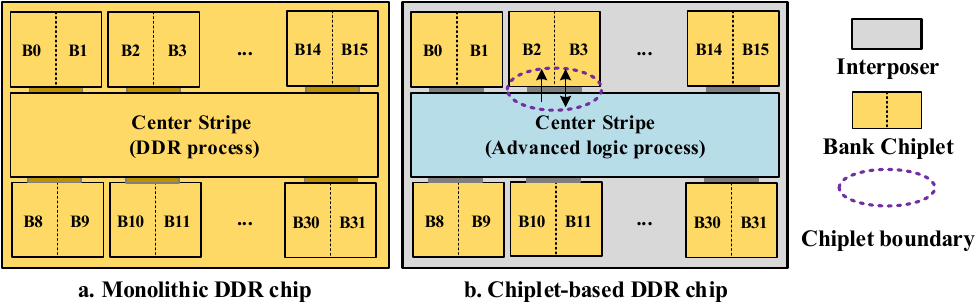}
    \caption{Chiplet DRAM.}
    \label{fig:chipletdram}
\end{figure}

This trend has also gained traction in DRAM design. Instead of a conventional all-DRAM-process design (\Cref{fig:chipletdram}a), recently introduced chiplet-based DRAM architectures \cite{ChipletTVLSI} separate the conventional DRAM array banks, fabricated using conventional DRAM processes, from the center stripe that includes signal driver, control and interface logic, which is relocated to a dedicated chiplet built with a faster logic process (\Cref{fig:chipletdram}b). This separation allows each component to benefit from its optimal fabrication technology. The DRAM process is optimized for high-density, low-leakage capacitive cells, while the logic process enables faster and more efficient control circuitry. Since the center stripe contributes significantly to critical timing parameters due to long signal propagation paths, relocating it to a logic chiplet can actually improve timing performance, and
even after accounting for die-to-die connections, buffers, and other chiplet-supporting components, the overall area is reduced by 8\% due to the area efficiency of advanced logic technology in the center stripe. A wire pitch of 4.7~µm has been shown to safely maintain signal integrity for supporting two orders of magnitude higher bandwidth than demands of DRAM systems, demonstrating the feasibility of this approach even for dense memory architectures.
In addition, this design enables the integration of multiple DRAM array chiplets on a single interposer. By sharing I/O pins and peripheral logic blocks, it reduces packaging cost and simplifies system integration when building DDR DIMM modules.  
Multi Chip on One Interposer (MCoOI) allows very high-density interconnect among chiplets.This enables a transformative new PIM capability: (1) we can still fully utilize the high internal DRAM bandwidth across the bank interfaces to overcome the memory bandwidth bottleneck in conventional accelerators; (2) meaning that we can take advantage of the highly optimized DRAM technology to provide dense and massive storage capacity for the large weight matrices; (3) while the processing units and DRAM peripheral logic can be fabricated on a separate die with logic process optimized for delay, and smaller area overhead, so that we can incorporate more aggressive compute units (such as systolic arrays) and large SRAM buffers to overcome the inefficiency of computing GEMM and supporting data reuse. 

%% file: 4-architecture.tex
\section{\nname{} Architecture} \label{sec:Architecture}

\begin{figure*}
    \centering
    \includegraphics[width=0.9\textwidth]{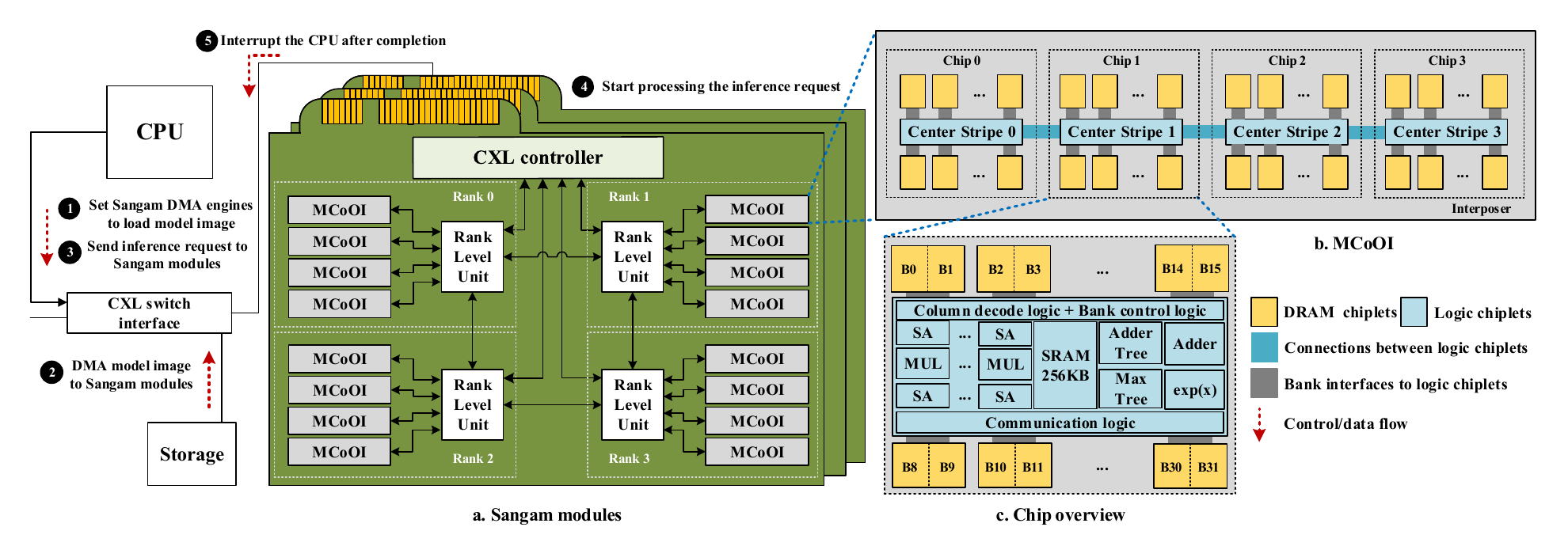}
    \caption{System Integration of the Sangam modules and organization. SA: systolic array. MUL: SIMD multiplier. MCoOI: multichip on one interposer
    }
    \label{fig:Sangam_system}
\end{figure*}

\subsection{System Overview and Integration}

\nname{} modules are CXL Type 3 devices that connect through the PCIe 6.0 standard.
Each module is connected to a host PC and other \nname{} modules using the CXL interconnect as shown in \Cref{fig:Sangam_system}.
State-of-the-art CXL switches can support up to 144 PCIe 6.0 lanes\cite{BroadcomPEX89144}, enabling a single switch to connect a maximum of 16 \nname{} devices (considering x8 lanes per device), leaving a x16 PCIe connection to the rest of the CXL interconnect or the host.

\nname{} modules have a discrete memory architecture\cite{nvidia2017v100,amd2023mi200}, separating the \nname{} module memory space from the CPU memory space.
The modules have their own DMA engines for data movement to and from the devices which are managed by the \nname{} driver.
Inter-module communication is done through peer-to-peer PCIe transactions \cite{CENT}. Having a separate memory space also allows the data mapping to Sangam to be different than conventional DDR, so that the various matrices are kept together in specific banks.

\subsection{Execution flow}
\Cref{fig:Sangam_system} shows a step-by-step illustration of the execution flow. The inference program for the model is statically compiled using the \nname{} memory configuration mapping. At module initialization time the \nname{} driver issues the command to DMA engines within the \nname{} modules to load the model checkpoint and the inference program(\circnum{1}). The DMA engines then load them into \nname{} modules(\circnum{2}). The previous steps only need to be done once before inference requests are received. Once the inference request is received, the host sends the inference request to the \nname{} modules using the driver interface(\circnum{3}). The input is DMAed to the modules, and upon completion, the modules start processing the inference request(\circnum{4}). Finally, the host is informed of inference completion using an interrupt(\circnum{5}).

\subsection{\nname{} module}
We maintain a memory hierarchy similar to conventional DRAM as proposed in \cite{ChipletTVLSI}, where DRAM memory bank chiplets and a center stripe chiplet are implemented on an interposer, forming a Multi-chip on One Interposer (MCoOI) as shown in \Cref{fig:Sangam_system}b. Each center stripe chiplet interfaces with all the banks within a conventional DRAM chip\footnote{Note that we don't have separate DRAM chip dies on the interposer. All the memory array banks of conventional DRAM chips, along with their associated peripheral and control logic, are all part of the yellow and blue chiplets illustrated in \Cref{fig:Sangam_system}b.} via chiplet-to-chiplet interconnects. A Sangam module can be configured to accommodate multiple such MCoOIs to form multiple ranks as illustrated in \Cref{fig:Sangam_system}a. Manufacturing and assembly yield study from \cite{ChipletTVLSI} recommends putting chiplets corresponding to four conventional DRAM chips onto a single interposer. As a baseline configuration for a DDR5 DRAM, we consider four chips per MCoOI, four MCoOIs per rank, and four such ranks make a Sangam module. Each MCoOI has 64 memory chiplets (each with two banks) and 4 center stripe logic chiplet (making it 17 chiplets for a conventional DRAM chip). Each center stripe is augmented with processing units and SRAM to facilitate computation and communication (see \Cref{subsec:chiplet_arch}).

\subsection{Logic Chiplet Architecture}
\label{subsec:chiplet_arch}

We integrate all the processing units required to support LLM inference operations, along with the conventional DDR5 DRAM central stripe logic, into a chiplet fabricated in 7nm logic technology, as shown in  \Cref{fig:Sangam_system}c. The logic chiplet is placed side‑by‑side with the memory bank chiplets on an interposer, enabling high‑bandwidth connections to all 32 DRAM banks per chip.  

Each bank is paired with an 8×8 FP16 systolic array and a 16‑lane SIMD FP16 multiplier. We add dedicated SIMD multiplier units to efficiently handle GEMV in the attention layer and other element-wise multiplication kernels, avoiding the severe under utilization and higher dynamic power that would result from mapping these operations onto the systolic array.
The systolic array width is fixed at 8 to match the DDR5 bank’s interface (128 bits), the base architecture has a height of 8. Each systolic array comprises FP16 multiply–accumulate (MAC) units, where multipliers have a two‑stage pipeline and adders have one pipeline stage. The 16‑lane SIMD FP16 multiplier is used for both GEMV kernels and element-wise operations such as in activation functions.
The row buffer interface of each bank is directly connected to the inputs of its systolic array and SIMD multiplier, enabling full bank‑parallel operation. All systolic arrays/multipliers across the 32 banks operate in lock-step at 400MHz, rate‑matched to the DDR5 bank bandwidth, with a \texttt{tCCDs} of 2.5ns, thereby fully utilizing the available all bank bandwidth.

\begin{figure*}[t]
    \centering
    \includegraphics[width=\textwidth]{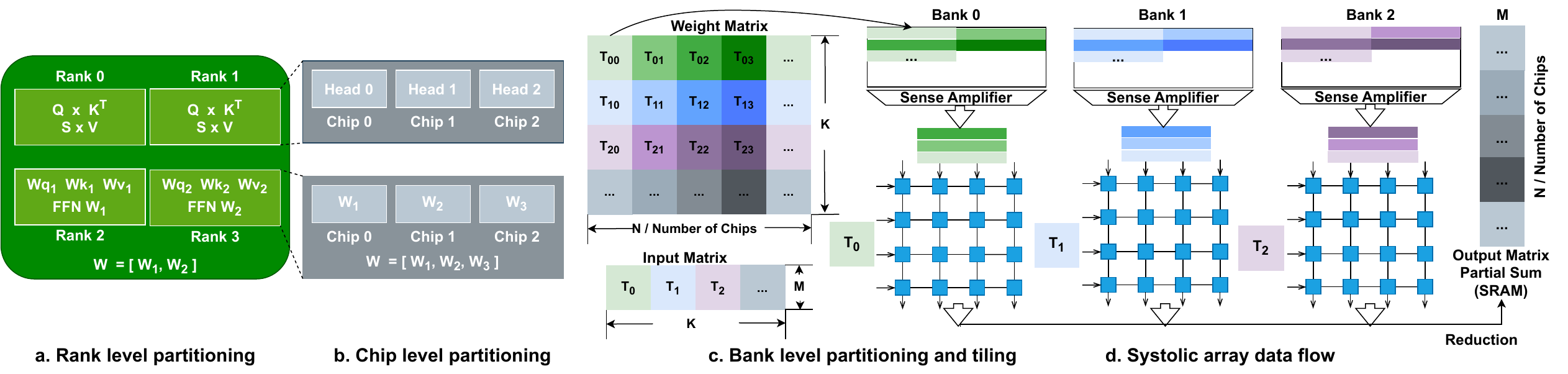}
    \caption{Hierarchical tensor mapping for flat GEMM execution across four levels: (a) rank: separating KV cache and weights, (b) chip: head-wise/column-wise partition, (c) bank: row-wise partition, and (d) systolic array: input-stationary. T: tiles of matrices that have $8*8=64$ FP16 values. Different colors (green, blue, purple, grey) indicate tiles mapped to different banks, while color intensity reflects the sequence in which tiles are read from the bank and streamed through the array.}
    \label{fig:mapping}
\end{figure*}

At the chip level, we employ eight 32‑to‑1 adder trees to gather and accumulate the outputs from systolic arrays or SIMD multipliers across all banks. The outputs of these adder trees are further accumulated with partial sums via a SIMD FP16 adder and then stored in a 256KiB on‑chip SRAM, which serves as a high‑bandwidth scratchpad for intermediate results. For reduction operations, the chiplet also integrates a 64‑to‑1 max‑reduction tree for argmax/softmax support and a 32‑lane SIMD FP16 exponential unit. Area breakdown of each of these components is discussed in \Cref{subsec:area_power}.

\subsection{Data Mapping and Partitioning}
\label{subsec:Data Mapping}

Flat GEMMs pose a challenge for conventional  systolic arrays, as the limited $M$ dimension underutilizes processing elements along the $M$ axis and reduces effective throughput.   \nname{}  addresses this by employing many small systolic arrays (e.g., $8\times 8$) distributed across the chips. 
To fully leverage this hardware organization and explore \emph{tensor parallelism}~\cite{shoeybi_megatron-lm_2020}, we design a \emph{hierarchical tensor mapping and partitioning} scheme that maximizes parallelism, minimizes communication, and balances workloads across ranks, chips, banks, and systolic arrays.

\paragraph*{Rank-level partitioning}
We disaggregate the model weights and the KV cache into different ranks. In the proposed \nname{} module, two ranks are dedicated to attention computation (where KV cache is allocated) referred as ``kv\_ranks" and two ranks to projection computation (where all the weight tensors corresponding to model are allocated) referred as ``wt\_ranks". Since the KV cache grows dynamically during inference, placing it in separate ranks from the weights both simplifies memory allocation within each chip and bank, and improves scalability by decoupling their storage requirements \cite{distserve}.  Furthermore, since the KV cache is unique for each request in a batch of requests, we do a round-robin allocation of KV cache per batch to the ``kv\_ranks" for parallel processing across batches. Note that the ``wt\_ranks" and ``kv\_ranks" have homogeneous center stripe architecture with systolic arrays to support models with GQA (Grouped Query Attention). The systolic array could be clock-gated when not in use.

\paragraph*{Chip-level partitioning}
Within a rank, we seek to exploit the parallelism of all 16 chips while avoiding chip-to-chip data transfers. We employ \emph{head-wise partitioning} for attention kernels in the ``kv\_ranks", and \emph{column-wise partitioning} of the weight tensor for projection kernels in the ``wt\_ranks".  
For attention kernels, all operands associated with a given attention head reside within the same chip, eliminating the need for inter-chip communication.  
For projection GEMMs, the $N$ dimension of the weight tensor is typically large; column-wise partitioning along $N$ allows us to distribute the workload evenly across chips and fully utilize parallel resources. The input matrix is broadcast to all chips, and the rank-level unit concatenates each chip’s output along $N$ to form the final result.  
Let $N_c$ be the number of chips in the module; each chip stores $N/N_c$ columns of the original weight matrix, and its local GEMM becomes: $\mathbf{X} \times \mathbf{W}_1$ with shape $(M, K, N/N_c)$.

\paragraph*{Bank-level partitioning}
To fully utilize aggregate bank bandwidth and further reduce GEMM size in a balanced way, we partition $\mathbf{W}_1$ \emph{row-wise} across all banks.  
Since each bank’s systolic array processes an $8\times 8$ tile at a time, we allocate 8 consecutive $\mathbf{W}_1$ rows per bank before moving to the next in a round-robin fashion, ensuring that different sets of 8 rows are assigned to different banks. For typical LLM weights, $K$ is typically divisible by $N_b$, which ensures balanced workloads and eliminates tail effects among banks. This row-wise partitioning splits the reduction dimension $K$ across banks, so partial sums must be reduced using \emph{adder trees} at the chip level. Within each $8\times 8$ tile, weights are laid out sequentially in DRAM so that consecutive columns can be streamed efficiently during systolic execution. After this mapping, the GEMM $\mathbf{X}_1 \times \mathbf{W}_1$ executed by each bank’s systolic array has the shape: $(M, K/N_b, N/N_c)$.

\paragraph*{Systolic array mapping}
We consider three classical dataflow strategies: \emph{input stationary}, \emph{weight stationary}, and \emph{output stationary}, each with different data reuse characteristics and on-chip storage requirements.  
For a GEMM of shape $(M, K/N_b, N/N_c)$, each weight element is reused $M$ times (often small, as discussed in \Cref{subsec:GEMM_Shapes_and_Ai}), each input element is reused $N/N_c$ times, and each output element is reused $K/N_b$ times, which is typically much larger than $16$.  
Given these reuse ratios, we adopt an \emph{input stationary} dataflow: tiles of the input matrix are preloaded into the systolic array to maximize reuse, while weights and partial sums stream through the array. Between input stationary and output stationary, the trade-off is between execution latency and SRAM capacity. When $K/N_b > N/N_c$, the output matrix is smaller than the input matrix, and its elements have higher reuse than input elements; however, preloading outputs into the array requires more SRAM for the larger input matrix. Our choice of input-stationary balances performance and hardware cost by reducing required output buffering while preserving full reuse of input tiles. The output tiles from the systolic arrays from all banks are reduced through the adder trees, and the (partial sum) output matrix is stored in the SRAM. 

\paragraph*{Hierarchical PIM architecture}
We envision a system of Sangam modules as a hierarchical network of logic units with tree topology, with the center stripe logic unit (chip-level unit) being the leaf node which connects to the rank-level unit on the PCB module as illustrated in \Cref{fig:Sangam_system}a. All the rank-level units connect to the CXL controller (channel-level unit). The CXL controller is connected to a CXL switch (root-level unit). All the parent units support reduction and aggregation operations to gather data from the children units. The kernels are mapped to the logic units based on the placement of different tensors in the memory (more in \Cref{subsec:HARMONI}). All the GEMM kernels gets mapped to the chip-level unit (enabling maximum tensor parallelism across chips in all ranks, and modules). All the reduction based kernels map to either the rank-level, channel-level, or root-level unit based on the Sangam system configuration. 


%% file: 5-methodology.tex
\section{Evaluation Methodology} \label{sec:Methodology}

\subsection{HARMONI}
\label{subsec:HARMONI}

\begin{figure}
    \centering
    \includegraphics[width=0.45\textwidth]{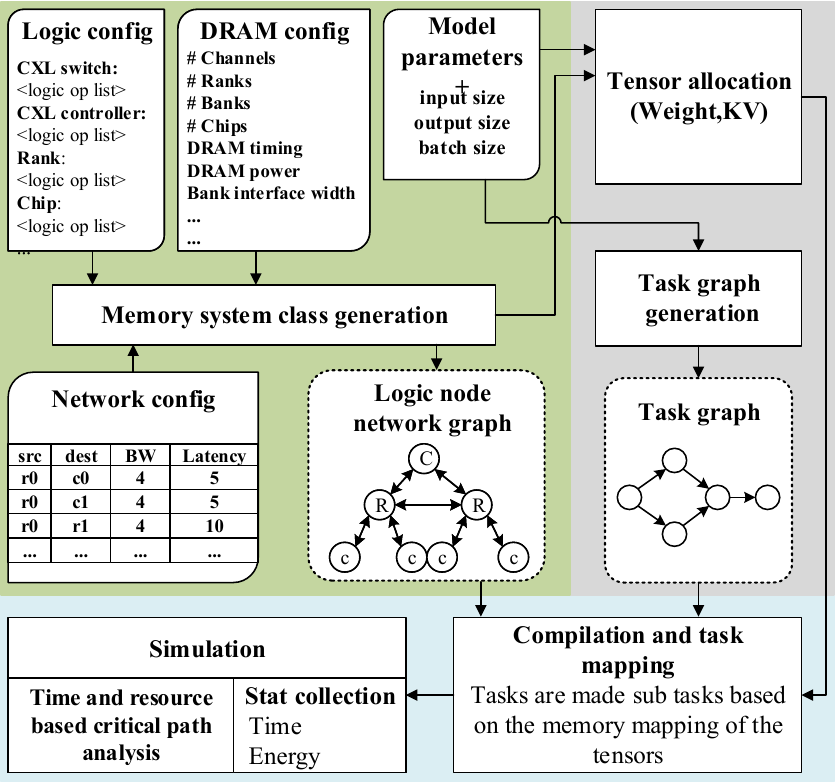}
    \caption{HARMONI modeling framework}
    \label{fig:framework}
\end{figure}

We developed a custom performance modeling framework \namef{} (\textbf{H}ierarchical \textbf{AR}chitecture \textbf{MO}deling for \textbf{N}ear/\textbf{I}n Memory Computing) that facilitates custom memory allocation (allowing us to determine the location of each tensor and determine appropriate memory access latency), inference program representation (acting as intermediate representation/execution trace in the form of a task graph), and task mapping to the logic nodes (mimicing program compilation and task scheduling). This framework enables a detailed analysis of LLM inference performance for different memory configurations and logic unit network. 
\Cref{fig:framework} illustrates the flow of the framework.

\subsubsection{Phase I - Setup}
\textbf{Memory System Generation.}
We define the memory system using a DRAM configuration, logic configuration, and network configuration.
The DRAM configuration defines the number of channels, ranks, banks, chips, bank\_interface width, DRAM timing parameters 
and DRAM current parameters 
, which is extracted from DRAMsim3 \cite{dramsim3}, Ramulator2.0 \cite{ramulator}, and memory vendors' datasheets for different DDR configurations. The logic configuration defines the supported operations at the different levels of logic unit network (chip, rank, channel, root). The network configuration defines the connection among nodes, e.g., bandwidth and link latency. This generates a memory system that is a directed graph, with the nodes as logic units, and edges define the interconnect between the logic units. Each logic unit is identified with an ID (i.e. the channel, rank, chip number). 

\subsubsection{Phase II - LLM inference program}
\textbf{Memory Allocation for Tensors.}
We use model configuration with all the hyper paramaters like \#heads, \#layers, dimensions, and create a dictionary of model weights ($Wqkv$, $Wo$, $Wup$, etc.). Each weight is defined by a custom tensor object 
that has fields such as shape, size, address\_offset. These tensors are allocated to the DRAM using the memory system. The allocation determines address offset of each tensor. The address offset is  a DRAM address following \textit{RoChRaCoBgBa} interleaving. Since we adopt all bank mode with lockstep execution on all chips within a rank, the granularity of each  memory access is $\#chips \times \#banks \times bank\_interface\_width$ (8KB for the base Sangam configuration). KV cache tensors can also be pre-allocated based on the batch size and maximum context length (batch-wise round-robin allocation to the ``kv\_ranks"). 
 
\textbf{Task Graph Creation.}
This step simulates a high-level program trace for LLM inference. Each kernel (e.g., QKV projection, activation, head-wise attention) is modeled as a node in a directed task graph, and edges capture data dependencies and tensor movement. Each node is annotated with metadata: kernel type, model phase (prefill or decode), input/output tensor accesses, etc. We validated the generated task graph with various LLM implementations on CPU and GPU by verifying the sequence of kernels and input, output tensor shapes. These metadata in the nodes helps with task mapping to logic nodes and statistics collection. We implemented QKV projection as a fused QKV projection to increase the embedding vector reuse, we also fused softmax with score calculation in the head-wise attention kernel (to reduce the number of memory accesses, thereby improving the performance).

\subsubsection{Phase III - Compilation and Scheduling}
\textbf{Task Mapping to Logic Nodes.}
In this compilation phase, each task node in the task graph is mapped to a logic unit in the memory system network. The mapping is determined by two factors: First, identifying the logic unit ID matching the tensor partitions (i.e., memory chip IDs where the input tensors corresponding to the task are allocated). Second, verifying that the logic unit supports the task kernel. The mapping algorithm minimizes tensor movement across chiplets (by using weight and KV tensor stationary strategy), so only the intermediate output tensors move on the logic node network (i.e., typically the hidden dimension - 8KB for LLaMA 2-7B model with FP16/BF16). If a kernel requires data from multiple DRAM partitions (e.g., different channels or ranks), it is split into sub-tasks, and an aggregation node is introduced on the nearest common parent logic node. The task graph is updated accordingly to reflect these decompositions. 

\subsubsection{Phase IV - Simulation and Statistics Collection}
Finally, once the task graph is updated with the mapping, each node in the task graph has an assigned logic unit. A simulation pass traverses the graph to estimate execution time for each kernel and communication time along graph edges. 
Since we have disaggregated ``wt\_rank" and ``kv\_rank", only one type of rank will be active at a time, so we use the idle time per rank to issue refresh commands. Once the execution time is determined, we generate a \textit{network trace} based on the communication configuration to determine the communication cost for edges in the task graph. The final inference latency is computed by identifying the critical path across the compute and communication nodes. 

\begin{table}[t]
\centering
\setlength{\tabcolsep}{3pt} 
\renewcommand{\arraystretch}{0.7} 
\caption{Interconnect parameters}
\label{tab:sangam_links}
\begin{tabular}{lcccc} 
\toprule
\textbf{Connection} & \textbf{Bandwidth (GB/s)} &
\multicolumn{3}{c}{\textbf{Latency (ns)}} \\
\cmidrule(lr){3-5}
 & & \textbf{Link} & \textbf{Src port} & \textbf{Dst port} \\
\midrule
Switch $\rightarrow$ CXL Ctrl & $128/\#\text{modules}$ & 20 & 25 & 5 \\
CXL Ctrl $\rightarrow$ CXL Ctrl & 32 & 20 & 5 & 5 \\
Rank $\rightarrow$ CXL Ctrl & 32 & 20 & 5 & 5 \\
Rank $\rightarrow$ Rank & 32 & 20 & 5 & 5 \\
\bottomrule
\end{tabular}
\end{table}

\textbf{Communication Cost Estimation:}
The parameters used in Sangam communication cost estimation are shown in \Cref{tab:sangam_links}.
The compute units between Ranks, Ranks to CXL controller, and between CXL controller are connected using PCIe 6.0 interconnect, with each having a 32 GBps bandwidth (for S-4M-4R-16C-128). 
The bandwidth between the CXL controllers of the Sangam modules and CXL switch is modeled based on the methodology used in \cite{CENT} where the total bandwidth is divided amongst the attached Sangam modules.

\subsection{Area, Power, Yield, Cost Estimation}
\label{subsec:area_power}
\textbf{Area.} We have designed all the processing units in Verilog and synthesized them using the Synopsys Design Compiler with a 14nm FinFET logic technology node. For modeling SRAM buffers, we employed FN-CACTI~\cite{ravipati_fn-cacti_2022} with the 14nm FinFET technology node. To account for the scaling effects from the 14nm logic process to a 7nm logic process, we scaled the area overhead of both the processing units and SRAM buffers based on the scaling factors outlined in~\cite{stillmaker_scaling_2017}. To estimate the area of the center stripe and bank of a DDR5 die, we used a 10nm 32Gb DDR5 die photo from~\cite{choi_132_2024}.
We treated the DDR5 die center stripe as equivalent to a 16\,nm logic process, and then scaled it down to 7\,nm~\cite{ChipletTVLSI}.

\textbf{Power.} We used the Micron DRAM Power Calculator~\cite{noauthor_dram_nodate} to model all the DRAM operation energy. We model the energy of bringing data from the cell array into our systolic array interface by accounting for both activation and read operations. Activation energy is calculated using IDD0 currents defined by JEDEC\cite{noauthor_ddr5_nodate} that capture the cost of opening (and closing) a row, which includes wordline driving, charge sharing, and latching all cells into the bank-level sense amplifiers. After activation, the data reside in the row buffer, and subsequent read commands only incur the column-path energy of selecting a slice from the row buffer. To quantify this cost, we again use the JEDEC-defined IDD4R\cite{noauthor_ddr5_nodate} currents for read operations. And since our systolic arrays directly connect the bank-level sense amplifiers and bypass the internal I/O drivers, we attribute only the portion of the read energy associated with local control and column select lines. Following prior breakdowns of DRAM column energy\cite{ha_understanding_2018}, we account for approximately 34\% of the read energy for this path, including control wires, CSL power, and global column address. The current specifications were gathered from the Ramulator2 simulator~\cite{luo_ramulator_2023}.
For GPU power consumption, we approximated the average power of the NVIDIA H100 SXM using 80\% of its thermal design power (TDP)~\cite{noauthor_nvidia_nodate}, in accordance with experimental observations by~\cite{CENT}.

\textbf{Yield \& Cost.} We adopt a yield estimation methodology consistent with that described in~\cite{ChipletTVLSI}. Specifically, the overall MCoOI yield is modeled as the product of three factors: (1) interposer yield, (2) DRAM/logic chiplet die yield, and (3) chiplet-to-interposer bonding yield. Using the approach in~\cite{kuo_overview_1999} for die yield,~\cite{graening_chiplets_2023} for the bonding yield, and the same wafer parameter assumptions as in~\cite{ChipletTVLSI}, we obtain an interposer yield of 94\%, a chiplet die yield of 97\%, and an overall effective yield of 90\%. For cost estimation, we adopt the same analysis framework as~\cite{ChipletTVLSI}, which incorporates four major components: (1) wafer and die cost, (2) DIMM packaging and assembly cost, (3) interposer cost, and (4) bonding cost. The parameter assumptions are aligned with prior works~\cite{graening_chiplets_2023, ChipletTVLSI, feng_chiplet_2022}.
Based on these assumptions, we estimate the cost per MCoOI to be \$3.85. At the module level, the total manufacturing cost of a \nname{} module is estimated to be \$61.99.

We extend our cost model to the full H100 module, which includes the GPU die, HBM, interposer, and assembly, with the cost dominated by HBM. We use the GH100 die area of approximately $814~\text{mm}^2$~\cite{noauthor_nvidia_2022} and assume advanced-package assembly at \$500 with a 97\% yield~\cite{noauthor_semiconductor_nodate}. For memory, we adopt a cost of \$110/GB for HBM2e~\cite{morgan_he_2024}, treating this as a conservative lower bound for HBM3. The assumptions for logic wafer cost and interposer cost follow those used in the \nname{} estimation. Based on these inputs, we estimate the GPU die cost to be approximately \$486, the interposer cost to be \$408, and the cost of six 16\,GB HBM stacks (94\,GB total) to be about \$1,760 per stack. Altogether, these values yield a total estimated cost of \$12,324 per H100 module.

\subsection{System configuration}

\Cref{tab:system_config} shows the different system configurations used for evaluation. \cite{CXL-PNM} suggests that a DDR5-based CXL module can support up to 512 GB, which is much greater than any reasonable HBM configuration. We compare Sangam to H100 GPU with 94 GB HBM, and a prior PIM-only work CENT \cite{CENT}. We used five different Sangam configurations as shown in \Cref{tab:system_config}. Design D1, and D2 are base configurations with 16 chips per rank. We consider D3, D4, D5 with 8 chips per rank to study the scaling of Sangam system.
The GPU experiments use \textit{vLLM} to run the LLaMA 2-7B model on H100(s). We use \namef{} and CENT's modeling framework to evaluate Sangam's and CENT's performance, respectively.

\begin{table}[t]
\centering
\caption{System configuration summary. \\Peak bandwidth for H100 refers to external memory; \\for CENT and Sangam (PIM), it refers to internal memory.}
\label{tab:system_config}

\resizebox{\columnwidth}{!}{%
\begin{tabular}{|l|c|c|c|c|}
\hline
\textbf{System} & 
\begin{tabular}[c]{@{}c@{}}Peak BW\\(TB/s)\end{tabular} & 
\begin{tabular}[c]{@{}c@{}}Peak GEMM\\(TFLOPS)\end{tabular} & 
\begin{tabular}[c]{@{}c@{}}Peak SIMD\\(TFLOPS)\end{tabular} & 
\begin{tabular}[c]{@{}c@{}}Capacity\\(GB)\end{tabular} \\
\hline
H100           & 3.35   & 1980  & --    & 94  \\
H100-2       & 6.7    & 3960  & --    & 188 \\
CENT-8         & 128.0  & --    & 64.0  & 128 \\
CENT-32        & 512.0  & --    & 256.0 & 512 \\
\hline
\multicolumn{5}{|c|}{\textbf{Sangam Configurations}} \\ \hline
S-4M-4R-16C-128 (D1) & 51.2 & 409.6 & 25.6 & 128 \\
S-8M-4R-16C-256 (D2) & 102.4 & 819.2 & 51.2 & 256 \\
S-8M-4R-8C-128 (D3) & 51.2 & 409.6 & 25.6 & 128 \\
S-8M-8R-8C-256 (D4) & 102.4 & 819.2 & 51.2 & 256 \\
S-16M-8R-8C-512 (D5) & 204.8 & 1638.4 & 102.4 & 512 \\
\hline
\end{tabular}
}

\vspace{3pt}
\parbox{\columnwidth}{\footnotesize
\textit{Note:} Sangam labels follow the format \texttt{S-4M-4R-16C-128GB}, which represents (Sangam)–(4 modules)–(4 ranks/module)–(16 chips/module)–(128 GB total capacity).``D1” denotes Design 1, referenced in \Cref{sec:results}. CENT-8 is 8 CENT devices, and CENT-32 is 32 CENT devices.
}
\end{table}


%% file: 6-results.tex
\section{Results}
\label{sec:results}

Because the LLM inference performance vary considerably depending on input length, output length, and batch size, we highlight several cases in this section. The two smaller input/output configurations (32/64, 128/256) are chosen based on the average input and output lengths observed in conversational and code-completion use cases with the Alpaca and ShareGPT datasets, as used in \cite{Neupim}. The larger configurations consider longer inputs and outputs that are becoming more common \cite{distserve}. Note: The speed up mentioned in the text is geomean speedup unless specified otherwise.

\subsection{End-to-End (E2E) Inference Speedup}

\begin{figure}
\centering
\includegraphics[width=\columnwidth]{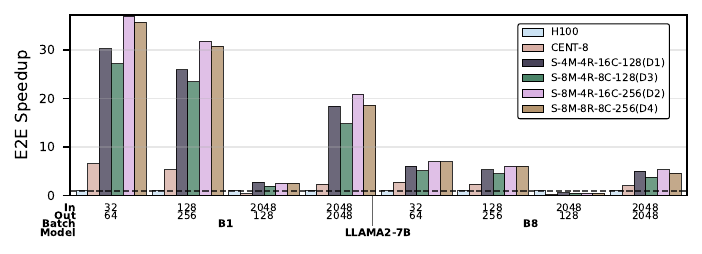}
\caption{E2E speedup over H100 for LLAMA2-7B}
\vspace*{-10pt}
\label{fig:e2e_speedup_llama2-7b}
\end{figure}

\begin{figure}
\centering
\includegraphics[width=\columnwidth]{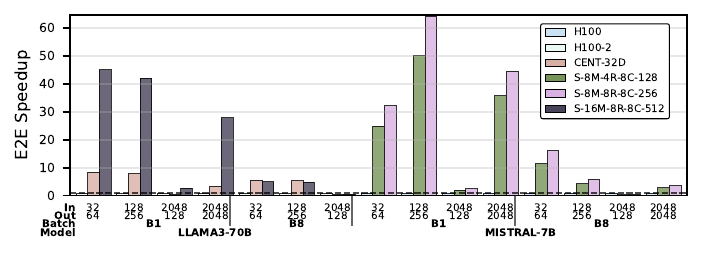}
\caption{E2E speedup for GQA supporting models}
\label{fig:e2e_speedup_llama3-70b}
\end{figure}

\Cref{fig:e2e_speedup_llama2-7b,fig:e2e_speedup_llama3-70b} compare end-to-end inference speedup normalized to H100(s) for LLaMA 2-7B, Mistral-7B, and LLaMA 3-70B on different Sangam and CENT configurations.

\textbf{LLaMA 2-7B.} We show the E2E speedup for Sangam (D1-4), and CENT-8 over H100. \textbf{Observation O1.} Sangam (D1-4) is almost always better than H100, (3.96$\times$) across the input/output/batch-size configurations shown. 
The only exception is for long input (2048) and short output (128 or less) for a larger batch (B8), where the Sangam (D1-4) E2E latency suffers a 0.55$\times$ slowdown vs.\ H100.
\textbf{O2.} Sangam (D1-4) performs better than CENT-8 by 3.49 $\times$. The speedup comes from Sangam's tensor-aware task mapping and accelerated GEMM execution using  systolic arrays. Hence, Sangam shows better performance than CENT on prefill-heavy workloads. For batch size  1, input 2048, and  output 128, Sangam D1 shows 2.76$\times$ geomean speedup on E2E latency whereas CENT-8 is slower than H100 by 0.38$\times$. \textbf{O3.} Inference with batch size 1 is faster by 3.3$\times$  compared to  batch size 8, which is 2.3$\times$ faster than H100. This is because single-batch inference is dominated by GEMVs, which are slow on H100.

\textbf{Mistral-7B, LLaMA3-70B.} \Cref{fig:e2e_speedup_llama3-70b} shows E2E speedup models with GQA implementation. Both Mistral-7B and LLaMA3-70B models have eight attention heads, we use Sangam D3, D4, D5 with 8 chips per rank configurations (1 head/chip). CENT does not have Mistral modeling, hence we only compare Sangam (D3, D4) with H100 for Mistral-7B. \textbf{O1.} D3 is 7.37$\times$, 2.2$\times$ better than H100 for B1, and B8 respectively, and D4 is 7.82$\times$, 1.96$\times$ for B1, and B8 respectively pertaining to more modules and higher parallelism. \textbf{O2.} Mistral-7B has higher performance benefit compared to LLaMA 2-7B because attention in Mistral-7B is flat GEMM (with M=8 due to GQA) benefiting from the systolic arrays. 
Since large models like LLaMA 3-70B does not fit on a single H100, we compare the performance of Sangam and CENT against 2 H100s (H100-2). We use 512 GB configurations of both Sangam (D5) and CENT-32 for this evaluation. \textbf{O3.} Compared to CENT-32, Sangam D5 performs better compared to H100-2 for batch size 1 (geomean speedup 4.2$\times$, showing least speedup of 2.5$\times$ for a large input to output ratio. \textbf{O4.} Sangam D5 experience a slight slowdown of 0.89$\times$ compared to CENT though faster than H100-2.

\subsection{Decode Throughput}
\begin{figure}
\centering
\includegraphics[width=\columnwidth]{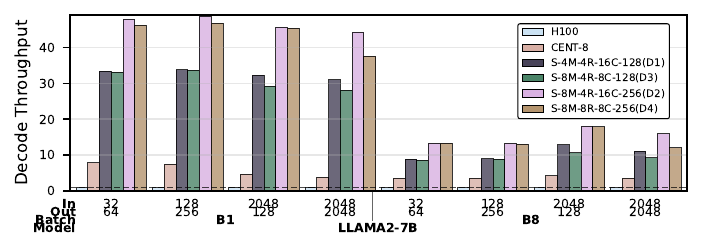}
\caption{Decode throughput normalized to H100}
\vspace*{-7pt}
\label{fig:decode_speedup}
\end{figure}

\begin{figure}
\centering
\includegraphics[width=\columnwidth]{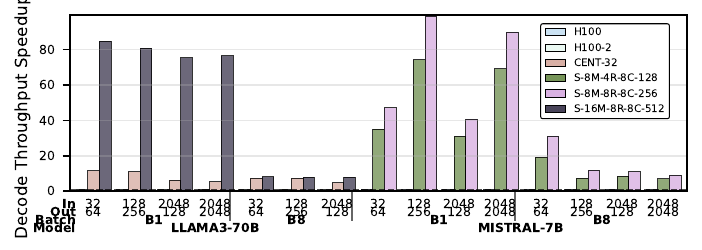}
\caption{Decode throughput comparison for GQA models}
\label{fig:decode_speedup_llama70b}
\end{figure}

\Cref{fig:decode_speedup,fig:decode_speedup_llama70b} show decode throughput for LLaMA 2-7B, Mistral-7B, and LLaMA 3-70B normalized to H100(s). 
\textbf{O1.} Sangam is always better than H100 for decode throughput by 10.48$\times$ for LLaMA 2-7B, since decode is memory-bound on GPU. This memory-boundedness is especially severe for larger inputs because of the larger KV caches.
\textbf{O2.} LLaMA 3-70B shows 4.08$\times$ higher decode throughput compared to CENT-32D because of the flat-GEMMs in the attention kernel of LLaMA 3-70B. \textbf{O3.} Mistral-7B shows 9.8$\times$ speedup compared to H100 due to higher memory bandwidth of Sangam modules (D3, D4) compared to H100, accelerated flat GEMMs due to GQA implementation. \textbf{O4.} Higher capacity D4 shows 1.3$\times$ higher speedup  than D3 for Mistral-7B.

\subsection{Crossover analysis considering TTFT}
Given the roofline analysis in \Cref{subsec:roofline} and OI discussion of different kernels in \Cref{tab:gemm_shapes_ai}, we expect PIM to perform poorly on compute intensive GEMMs, mainly in the prefill phase, compared to the GPU---impacting both TTFT and E2E latency.
In \Cref{fig:e2e_speedup_llama2-7b}, we see that the H100 beats all Sangam configurations for large input and small output (2048/128) for batch size of 8, and in \Cref{fig:TTFT_crossover}, which plots TTFT  as a function of input and batch size, we see that the H100 beats D1 for input lengths larger than 256 and a batch size of 1, and for inputs larger than 32 with a batch size of 8 (it is the product  that matters most). We also see that CENT  has significantly worse TTFT. One solution is to use the GPU for prefill (projections), perhaps only when the input length exceeds the TTFT crossover point. 

{\bf O1.} However, TTFT only matters if it hurts the user's perception of interactivity by making the interaction appear laggy. Otherwise, end-to-end query latency and throughput matter more to the user and for system efficiency, and PIM's advantages in decode throughput can make up for slower TTFT.  Different service level objectives (SLOs) for TTFT have been proposed.  In addition to plotting TTFT, \Cref{fig:TTFT_crossover} shows where Sangam's and CENT's TTFTs violate the SLO for TTFT targets of 0.5, 1.5, and 3.0 sec. (drawn from~\cite{layerkv, distserve}) as a function of output length.  {\bf O2.} Sangam has no violations when batch size is 1, for any input length we studied, while CENT suffers violations for shorter inputs.  Furthermore, even with batch size of 8, Sangam can accomodate fairly large inputs, up to 425 and 1129 for SLOs of 0.5 and 1.5 sec, and no violations up to 2048 for an SLO of 3 sec.  CENT again suffers violations for shorter inputs, i.e., 278, 678, and 1288.

\input{crossover} 

\begin{figure}
\centering
\includegraphics[width=0.95\columnwidth]{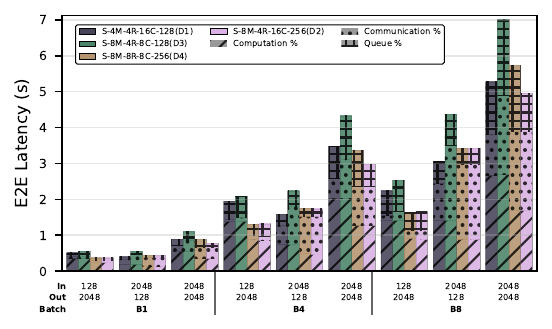}
\caption{E2E latency breakdown on LLaMA 2-7B}
\label{fig:e2e_breakdown}
\end{figure}

\subsection{Scaling study for Sangam on LLaMA 2-7B}
\label{sec:scaling}
\Cref{fig:e2e_breakdown} shows E2E latency breakdown in terms of time spent on computation, communication, and queueing delay for four different Sangam configurations (D1-4) for different in/out/batch sizes. Queueing delay here refers to the time spent by a task in the task graph waiting for computational resources. This leads to several observations:
\textbf{O1.} Queueing delay increases when reducing the number of chips and ranks (D1(21\%)$\rightarrow$D3(23\%)), (D4(19\%)$\rightarrow$D3(23\%)) since the tasks will be experiencing more resource contention.
\textbf{O2.} Increasing capacity improves the performance (D1,D3$\rightarrow$D2,D4) by reducing compute time, but the contribution of communication to E2E latency increases from 14.5\% to 28.5\%. Queueing delay reduces due to more resources (chips and ranks).
\textbf{O3.} Larger inputs lead to more communication overhead as the activations are large compared to the decode phase activations and embedding vectors.
\textbf{O4.} Increasing chips while decreasing  modules (D3$\rightarrow$D1) leads to reduction in compute, queueing time but communication increases, which is an overall win for the cases evaluated.
\textbf{O5.} Increasing the chips but decreasing the ranks (D4$\rightarrow$D2) improved performance without any significant change in the breakdown.
Observations are similar across all batch sizes and models.

\begin{figure}[t]
  \centering
  \begin{minipage}[b]{0.47\columnwidth}
    \includegraphics[width=\textwidth]{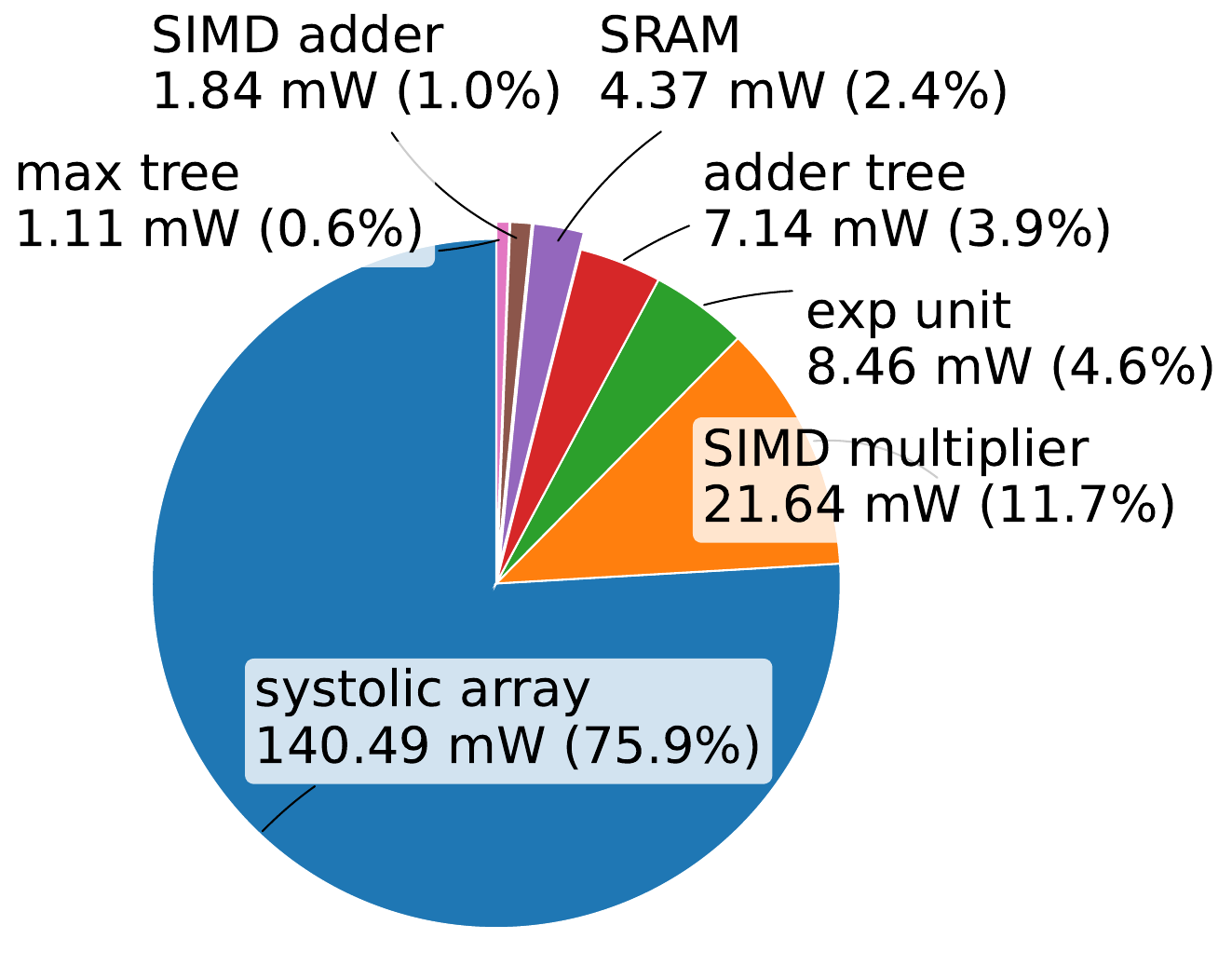}
    \caption{Center stripe chiplet PIM logic power breakdown}
\label{fig:chiplet_power}
  \end{minipage}
 \hspace{0.02\columnwidth}
    \begin{minipage}[b]{0.47\columnwidth}
    \includegraphics[width=\textwidth]{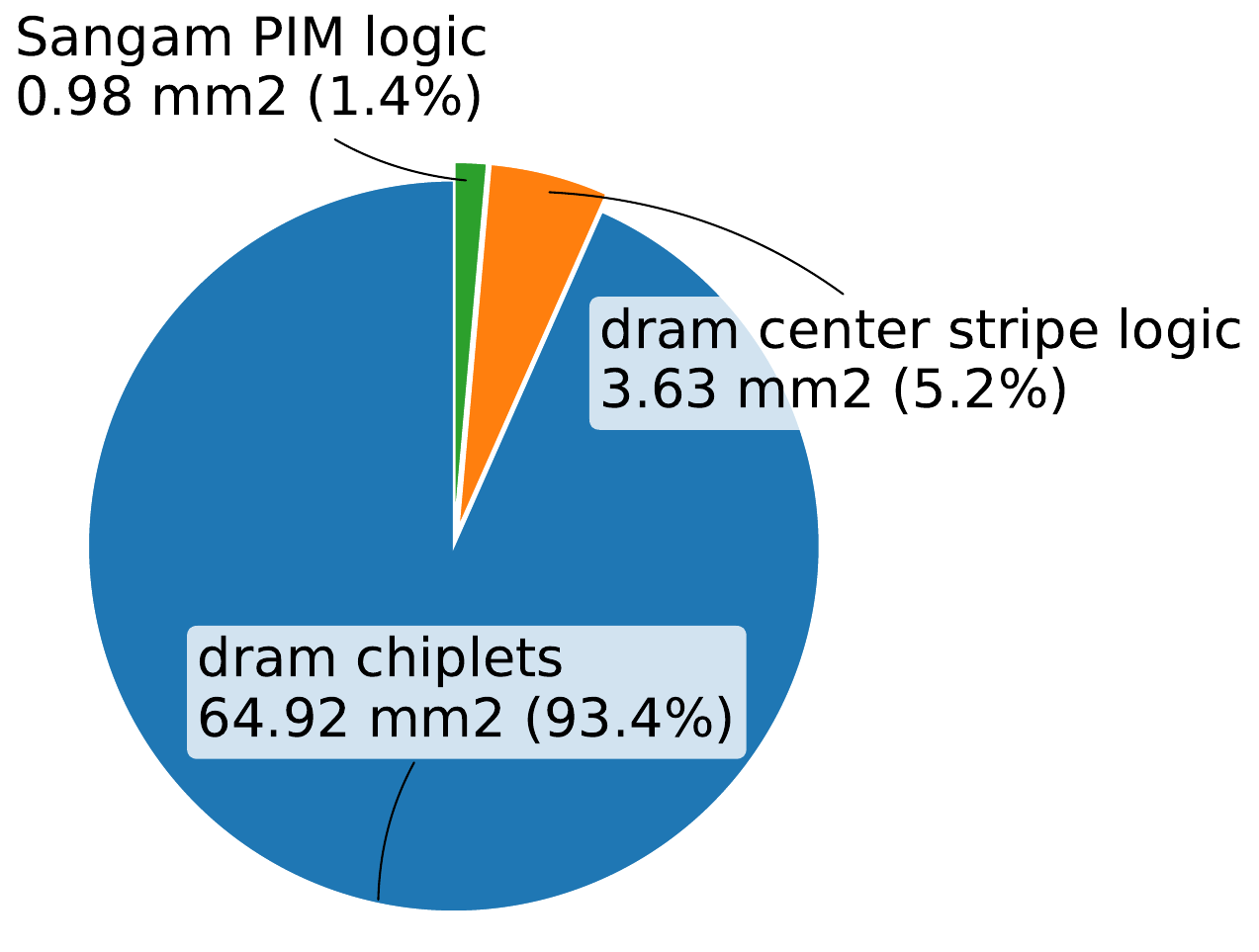}
    \caption{Chiplet area breakdown}
\label{fig:Chiplet_area_breakdown}
  \end{minipage}
\end{figure}

\subsection{Power, Area and Energy}
\label{subsec:power_area_energy}

\begin{figure}[t]
\centering
\includegraphics[width=\linewidth]{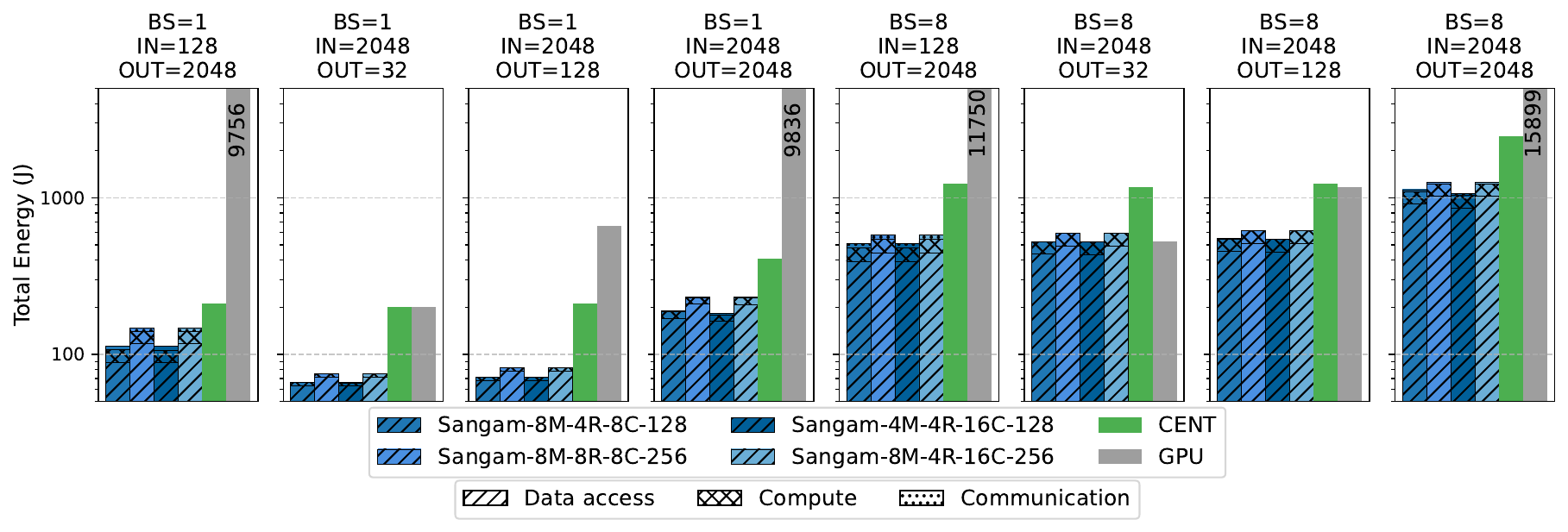}
\caption{Total energy consumption with breakdown of data access, computation, and communication}
\vspace*{-8pt}
\label{fig:total_energy}
\end{figure}

Figure~\ref{fig:chiplet_power} presents the power breakdown of the PIM logic functions within the center stripe chiplet. The systolic array dominates the logic power, followed by the SIMD multiplier, exponentiation unit, and smaller components such as the SRAM interface, adder tree, and max tree. Together, these units consume a total of approximately $185$~mW. Although this figure only reports the relative contributions of logic units, we note two important implications. First, given the total logic power of $185$~mW and the area of $0.98~\text{mm}^2$ allocated for these functions, the resulting power density is about $0.19~\text{W}/\text{mm}^2$, which is within the cooling limits typically sustained by commercial accelerators. Second, while the logic contributes a modest fraction of power, the dominant source of energy consumption arises from memory accesses.
(Figure~\ref{fig:total_energy} shows a more detailed analysis of energy breakdown, which shows that data movement significantly outweighs logic operation energy). Nevertheless, when combining the PIM logic power ($185$~mW) with the all-bank read power ($1.735$~W), the total per-chip power is about $1.92$~W. Since a Sangam module integrates $16$ such chips, the overall module-level power is approximately $30.7$~W, which remains within the power delivery constraints of a typical accelerator.

Figure~\ref{fig:Chiplet_area_breakdown} shows the area breakdown of one chip, which contains 16 DRAM chiplets and one logic chiplet. The logic chiplet consists of both the DRAM center stripe logic and the PIM logic. Based on this methodology, the two DRAM chiplets together occupy $93.4\%$ of the overall chip area, while the DRAM center stripe logic and the PIM logic occupy only $5.2\%$ and $1.4\%$ of the overall area, respectively. Thanks to the chiplet integration, we achieve significant area savings compared to a monolithic DRAM-PIM die.

\Cref{fig:total_energy} reports the total energy consumption of LLaMA2-7B inference across a range of input/output/batch size combinations on an H100 GPU, CENT, and four variants of Sangam. {\bf O1.} Except for the case of batch size 8, input size 2048, and output size 32, Sangam achieves order-of-magnitude energy reduction compared to GPU. Sangam also consistently achieves markedly lower energy consumption compared to CENT across all batch sizes and input/output lengths, although the magnitude of improvement varies with workload characteristics. {\bf O2.} Across all Sangam configurations, energy consumption is dominated by data access (DRAM activation and read), while computation and communication account for only a small fraction (around 5\%-20\%) of the total. This observation underscores that further optimization opportunities lie in reducing data movement. {\bf O3.} The total energy of the four Sangam variants is very close and  slightly higher for the configurations with more modules or ranks. This behavior is expected in a data-movement-dominated regime: the total amount of data accessed is similar, so increasing modules/chips increases parallelism, which improves latency/throughput rather than reducing energy;  the energy–performance trade-off is that more modules/ranks  buy lower latency or higher throughput at modestly higher energy.


%% file: crossover.tex
\begin{figure}[t]
\centering
    \includegraphics[width=\linewidth]{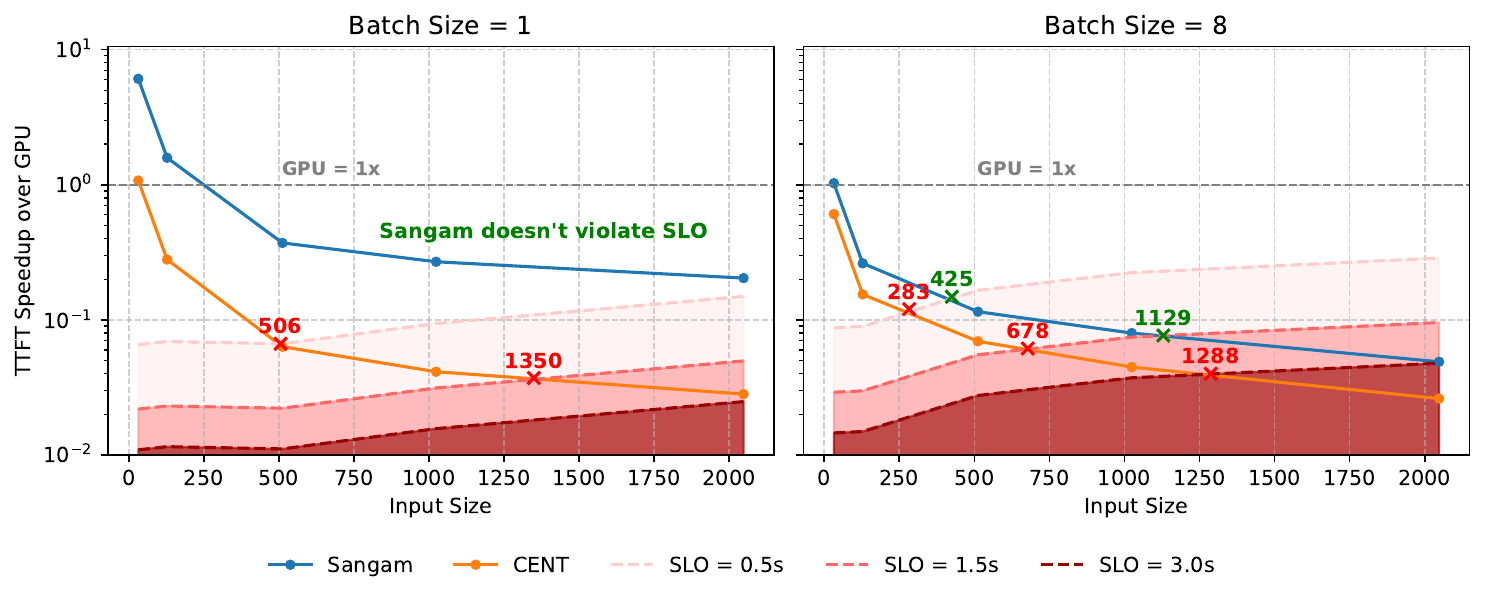}
    \caption{TTFT crossover as a function of input size. Since the plots are speedup over the GPU, we convert the SLO into a speedup threshold (against the GPU) for each batch size. SLO line represents minimum speedup needed to meet \{0.5, 1.5, 3\} sec.\ TTFT latency.}
    \label{fig:TTFT_crossover}

\end{figure}

%% file: 7-discussion.tex
\section{Related work}
\begin{figure}[t]
    \centering
\includegraphics[width=0.95\columnwidth]{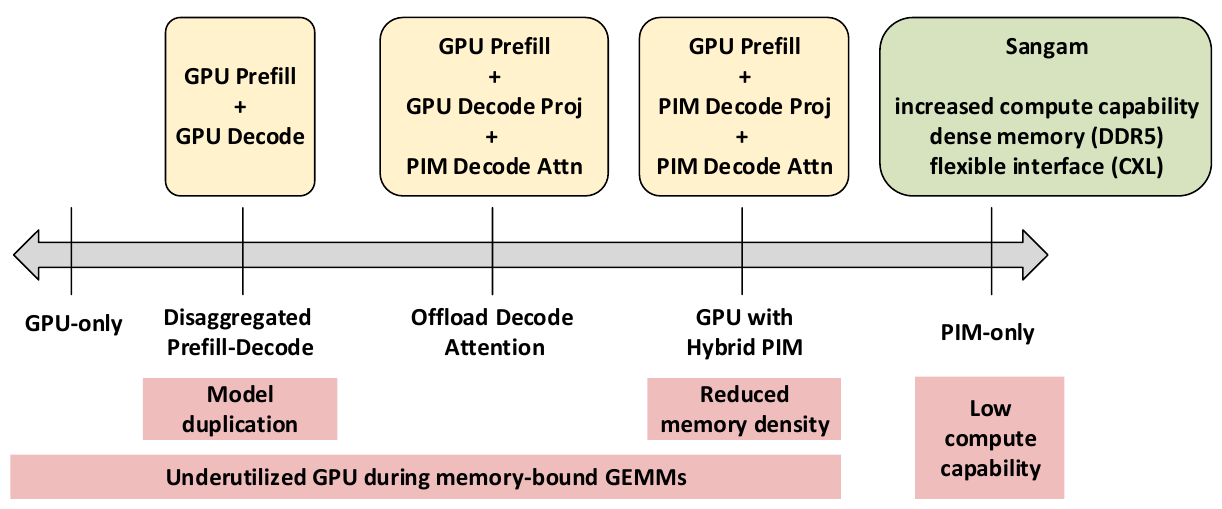}
    \caption{LLM Inference Acceleration Works}
\label{fig:Spectrum}
\end{figure}

Identifying the memory-intensive nature of LLM inference workloads, we observe that there is a spectrum of accelerator designs, as illustrated in \Cref{fig:Spectrum}, that have been proposed by the community. On one hand, there are {\bf GPU-only} solutions that leverage high-bandwidth HBMs to overcome the memory bandwidth limitation, however, HBMs are capacity-bound, resulting in multi-GPU setups or offloading model weights or KV cache to host CPU memory. Moreover, the under-utilization of GPUs during the decoding phase has motivated researchers to look for solutions to balance  prefill and decode latency. \cite{distserve} proposes to disaggregate  prefill  and decode on separate GPUs {\bf (GPU+GPU)} with different compute capabilities, however, this needs model weight duplication. To overcome the memory-bandwidth issue, many prior works propose hybrid solutions that attach a high throughput machine like GPU or NPU to a PIM/PNM solutions, e.g., \cite{attacc, IANUS, Neupim, PAPI, duplex, AiM}. Many related works in these hybrid {\bf (xPU+PIM)} settings either use HBM or GDDR which are limited in capacity. AttAcc \cite{attacc} proposes to attach HBM-PIM devices to GPU/xPU to offload only the attention kernels in the decode phase to PIM. AiMX \cite{AiM} prototyped by SK hynix leverages GDDR6-PIM (adding light-weight GEMV units per bank along with a small global SRAM buffer per die) to accelerate memory-bound GEMV computation in LLM inference. However, a na\"ive solution leaves the GPUs (or any AI accelerator) under-utilized on flat GEMMs. PAPI \cite{PAPI} proposed a dynamic scheduling strategy with hybrid PIM design: Attn-PIM for decode attention with low compute and FC-PIM for decode projection (mostly flat GEMMs) with high compute. However, adding logic in FC-PIM comes at the cost of reducing the number of banks or their capacity. Furthermore, logic in a DRAM process is costly in area and slow. Duplex \cite{duplex} proposes to increase the number of TSVs to enable more bandwidth at the logic die of the HBM (proposing a significant change to the architecture) to support GEMM operations. 
At the other end of the spectrum are {\bf PIM-only} solutions like CENT \cite{CENT} that integrates both in-memory PIM components at the DRAM bank level and near-memory processing elements at the CXL controller to accelerate LLM inference; however, due to DRAM technology limitations, it only supports GEMV but not GEMMs. Another CXL-based work using LPDDRX \cite{CXL-PNM} focuses on near-memory processing implemented at the CXL controller to support end-to-end LLM inference; however, PNM solutions expose lesser internal bandwidth than PIM, albeit cheaper. In contrast to these prior works, our approach, Sangam, distinguishes itself by introducing a chiplet-based DRAM-PIM architecture for end-to-end LLM acceleration, offering increased compute capability, higher memory capacity with DDR5 compared to HBM and GDDR, and a flexible CXL interface for attaching modules to GPUs and/or storage devices.
Beyond PIM and hybrid GPU-PIM approaches, other accelerator designs such as Cerebras \cite{cerbras}, Graphcore \cite{graphcore}, and TPUs \cite{ironwood_tpu_2025} are LLM accelerators designed to enhance inference performance by customizing on-chip hardware for Transformer workloads. However, these architectures largely overlook the fundamental bottleneck posed by the memory wall. In contrast, our work tackles this challenge from the ground up, focusing on mitigating the memory wall as a primary design constraint.

%% file: 8-conclusion.tex
\section{Conclusions}
Chiplet technology provides the best of both worlds: the ability to use the high density of DRAM for data storage, but the density and speed of logic for computation, while providing access to the massive internal bandwidth of the DRAM architecture. Sangam
provides a cost-effective, scalable,  memory-centric platform for LLM inference at scale, achieving 3.93, 4.22, 2.82$\times$ speedup in end-to-end query latency, 10.3, 9.5, 6.36$\times$ greater decoding throughput, and orders of magnitude energy savings compared to an H100 GPU for varying input size, output length, and batch size on LLaMA 2-7B, Mistral-7B, and LLaMA 3-70B, respectively.

%% file: 9-acknowledgement.tex
\section*{Acknowledgements}
This work was supported in part by PRISM, one of seven centers in JUMP 2.0, a Semiconductor Research Corporation (SRC) program sponsored by DARPA.